# Structural and Electronic Recovery Pathways of a Photoexcited Ultrathin VO$_2$ Film

Haidan Wen,[1,*] Lu Guo,[2] Eftihia Barnes,[2] June Hyuk Lee,[1] Donald A. Walko,[1] Richard D. Schaller,[3,4] Jarrett A. Moyer,[5] Rajiv Misra,[6] Yuelin Li,[1] Eric M. Dufresne,[1] Darrell G. Schlom,[7,8] Venkatraman Gopalan,[2,†] John. W. Freeland [1,§]

[1] *Advanced Photon Source, Argonne National Laboratory, Argonne, Illinois 60439, USA*

[2] *Department of Materials Science and Engineering, Pennsylvania State University, University Park, Pennsylvania 16802, USA*

[3] *Center for Nanoscale Materials, Argonne National Laboratory, Argonne, Illinois 60439, USA*

[4] *Department of Chemistry, Northwestern University, Evanston, Illinois 60208, USA*

[5] *Department of Physics, University of Illinoi at Urbana-Champaign, Urbana, Illinois 61801, USA*

[6] *Department of Physics, Pennsylvania State University, University Park, Pennsylvania 16802, USA*

[7] *Department of Materials Science and Engineering, Cornell University, Ithaca, New York 14853, USA*

[8] *Kavli Institute at Cornell for Nanoscale Science, Ithaca, New York 14853, USA*

## Abstract

The structural and electronic recovery pathways of a photoexcited ultrathin VO$_2$ film at nanosecond time scales have been studied using time-resolved x-ray diffraction and transient optical absorption techniques. The recovery pathways from the tetragonal metallic phase to the monoclinic insulating phase are highly dependent on the optical pump fluence. At pump fluences higher than the saturation fluence of 14.7 mJ/cm$^2$, we observed a transient structural state with lattice parameter larger than that of the tetragonal phase, which is decoupled from the metal-to-insulator phase transition. Subsequently, the photoexcited VO$_2$ film recovered to the ground state at characteristic times dependent upon the pump fluence, as a result of heat transport from the film to the substrate. We present a procedure to measure the time-resolved film temperature by correlating photoexcited and temperature-dependent x-ray diffraction measurements. A thermal transport model that incorporates changes of the thermal parameters across the phase transition reproduces the observed recovery dynamics. The optical excitation and fast recovery of ultrathin $VO_2$ films provides a practical method to reversibly switch between the monoclinic insulating and tetragonal metallic state at nanosecond time scales.

*wen@aps.anl.gov

†vxg8@psu.edu

§freeland@aps.anl.gov

# I. Introduction

Metal-to-insulator phase transitions (MIT) have been one of the central research frontiers in condensed matter physics [1–3]. The abrupt change in conductivity across the transition originates from strongly correlated interactions that cannot be explained by simple band theory. Excitation of the phase transition can be achieved by tuning temperature, pressure, chemical doping, strain, external fields or light. Among these excitation schemes, ultrafast optical excitation not only offers a unique approach to understanding a phase transition arising from collective electronic and structural rearrangements in the time domain [4–6], but also provides a means to manipulate properties of materials across the phase transition at ultrafast time scales [7].

Vanadium dioxide ($VO_2$) is an intensively studied example of a material which undergoes an MIT [1]. Although its transition mechanism has not been fully understood, it holds the promise of future MIT-based electronics, since its transition temperature is tunable in the vicinity of room temperature [8,9]. When the temperature of $VO_2$ decreases across the transition temperature, a substantial electronic structure change [10,11] yields a reduction in the $VO_2$ conductivity by several orders of magnitude [12]. Correspondingly, the tetragonal lattice structure in space group $P4_2/mnm$ is transformed to the monoclinic lattice structure in space group $P2_1/c$. The decrease of the structural symmetry is a result of dimerization of vanadium atoms that twist in pairs along the tetragonal *c*-axis [13]. The relationship between electronic and structural phase transitions has been at the heart of the debate regarding the phase transition mechanism in $VO_2$. The phase transition can be a result of a complex interplay of strong electron-electron interactions representing Mott physics [14] and the structurally-driven phase transition representing Peierls instability [15], in which the changes of orbital occupation have also been observed [16]. Using pump-probe techniques, the measurement of the time scales that are associated with electronic and structural phase transitions shed light on the transition mechanism. The photoinduced phase transitions of $VO_2$ were studied by optical [17–20] and terahertz [21–25] probes, primarily sensitive to the electronic properties, and by electron [26] and x-ray diffraction [27–29], directly measuring the lattice structure. These extensive works have been focused on the transition pathway from the insulating to metallic state. However, the dynamics of the reverse process, namely, the transition from metallic to insulating state that occurs over multiple time scales [30], is less explored.

In this article, we have investigated the structural and electronic recovery dynamics of a photoexcited ultrathin $VO_2$ film using time-resolved x-ray and optical probes at ns time scales, with an emphasis on the structural aspect of the recovery and its correlation with the electronic recovery process. Depending on the optical fluence, we find that the structural recovery can deviate from a thermal cooling pathway. At high pump fluence, a non-thermally induced transient structural state was identified through a measurement of the film temperature at ultrafast time scales. This transient structural state was decoupled from the MIT and recovered faster than the thermally allowed rate, corresponding to a relaxation of a photoinduced bond-softening state. The subsequent recovery at longer times was connected to thermal transport from the film to the substrate. We incorporated the change of thermal properties, including specific heat and thermal conductance of $VO_2$ during the phase transition, into a thermal transport model that reproduced the observed recovery dynamics. Understanding the thermal recovery process of $VO_2$ has been increasingly important since the voltage-induced phase transition in $VO_2$ in a two-terminal device geometry has been shown to be largely due to Joule heating [31].

From an application point of view, the ability to control the reversible switching between metallic and insulating states is crucial to the potential applications of $VO_2$ in ultrafast electronic devices and future information processing hardware [7]. Although the phase transition from the insulating to metallic state can be triggered by various stimuli, there is yet no active method to switch VO$_2$ from the metallic phase back to the insulating phase at ultrafast time scales. An ultrathin VO$_2$ film grown by molecular beam epitaxy provides an ideal system to allow fast cooling of the film due to efficient heat transfer from its crystalline layer through the interface to the substrate, leading to ns switching of VO$_2$ from the metallic to insulating state. Utilizing the ultrafast excitation and relaxation, we demonstrate a simple, effective method to allow fast switching between the excited metallic state and the insulating ground state up to GHz rates.

## II. Experiments and results

### a. Sample preparation and characterization of VO$_2$ thin films at thermal equilibrium

A 13-nm thick VO$_2$ film was grown on (100) TiO$_2$ substrates using reactive molecular beam epitaxy [32,33]. Electron transport measurements showed that the resistivity of the film was reduced by four orders of magnitude when the film temperature increased across the transition temperature, to accompany the structural phase transition shown schematically in Fig.1.

The lattice structure of VO$_2$ at thermal equilibrium was characterized by diffraction using 10 keV x-rays at the Advanced Photon Source (APS). The Bragg peaks were indexed according to the tetragonal lattice. A reciprocal space map around the 211 Bragg peak showed the VO$_2$ film was not commensurate with the substrate, due to large in-plane lattice mismatch of 3.6% along the tetragonal *c*-axis. The structural changes across the MIT at thermal equilibrium were monitored by measuring the 400 (monoclinic 040) Bragg peak of the VO$_2$ film [Fig. 2(a)], sensitive to the out-of-plane lattice parameter. The Bragg peak position and intensity were extracted and plotted as a function of temperature [black curves in Fig. 2(c)]. Below the phase transition temperature, the $VO_2$ lattice exhibited a monoclinic structure with an out-of-plane lattice parameter of 4.527 Å. When the film temperature was elevated across the transition temperature, the $VO_2$ film underwent the monoclinic-to-tetragonal structural phase transition where the tilted dimers of vanadium atoms along the *c*-axis straightened to form a tetragonal lattice structure [13,34]. As a result of the structural phase transition, the measured out-of-plane lattice parameter increased to 4.539 Å, corresponding to a 0.1° shift of the 400 Bragg peak to lower angle. The Bragg peak intensity also varied as a function of temperature. It decreased to the lowest value at 342 K, accompanied by a broadening of the Bragg peak, indicating disordering of the (100) lattice planes during the de-dimerization of the vanadium atoms. The loss of the diffraction intensity is associated with the increase in the number of phase boundaries as seen by microscopic probes [35,36], and possibly an intermediate structural state such as the $M_2$ phase [37] and monoclinic metallic state [18]. Above the transition temperature, the diffracted intensity increased to a higher value compared to that of the monoclinic phase, consistent with higher symmetry of the tetragonal lattice structure.

In comparing these results to the electrical transport measurements, we find the reduction of the film resistivity occurs in the same temperature range as the changes of the Bragg peak position, while the variation of the intensity of the Bragg peak spans a wider temperature range. Therefore, the electronic properties are highly correlated with the lattice structure but not sensitive to the ordering of the lattice.

### b. Time-resolved x-ray diffraction measurement

To study the structural recovery dynamics of a photoexcited $VO_2$ film, we measured the out-of-plane lattice parameter after photoexcitation by time-resolved x-ray diffraction at the 7ID-C station of the APS [38]. An 800 nm, 60 fs, laser pulse from a Ti:sapphire laser system at a repetition rate of 1 kHz was focused to a 1-mm full-width half-maximum (FWHM) spot on the $VO_2/TiO_2$ sample mounted on a temperature-controlled stage. We report the magnitude of the laser excitation in terms of the incident fluence. Since the absorption depth of 800-nm light of $VO_2$ is about 100 nm [27,39], the 13-nm thick $VO_2$ film was uniformly excited. The homogenous excitation circumvented any complications arising from the phase front propagating along the laser penetration direction. A 100-ps x-ray pulse was monochromatized to 10 keV from the storage ring and focused to a 50 μm FWHM spot, coincident with the optical pump pulse on the sample. The optical pulse was electronically synchronized to the x-ray pulse with adjustable delay.

### c. Calibration of the film temperature at 100 ps after optical excitation

In order to understand the recovery dynamics, it is crucial to identify the initial photoexcited state. Upon photoexcitation, the $VO_2$ film undergoes a photoinduced phase transition in the first several hundred fs, followed by electron-phonon interactions in which a thermally activated phase transition occurs in ps time scales [26,27]. The complete transformation of the excited $VO_2$ from the insulating to metallic phase can take a hundred ps as a result of the dynamical growth and percolation of the metallic domains [22]. Therefore, the photoinduced effect beyond 100 ps mainly increases the film temperature. We thus use the film temperature to characterize the initial state of the $VO_2$ film before its recovery, with exceptions at high pump fluences. Here, we describe a procedure of measuring the film temperature at 100 ps after photoexcitation.

We first recorded the 400 Bragg peaks of the $VO_2$ film as a function of pump fluence at a fixed time delay of 100 ps between the optical pump and x-ray probe pulses [Fig. 2(b)]. We then compared the Bragg peaks as a result of the photoexcitation to the Bragg peaks at various film temperatures. If the photoexcited and thermally excited Bragg peak positions and intensities matched, the film 100 ps after photoexcitation was designated by the temperature of the corresponding thermally excited film [Fig. 2(c)]. Therefore, the film temperature 100 ps after excitation in the fluence range of 7 to 14.7 mJ/cm$^2$ can be accurately calibrated due to the well-matched characteristics of the Bragg peaks. For example, the film temperature at 100 ps after photoexcitation was estimated to be 338 K with the pump fluence of 10 mJ/cm$^2$ as shown in Fig. 2(c). Beyond 14.7 mJ/cm$^2$, the Bragg peaks of optically and thermally excited VO$_2$ did not match. But the film temperature still can be estimated after correcting for the change of optical reflectivity and heat capacity across the phase transition. Since no further structural phase transition occurs in tetragonal VO$_2$, the change of the film temperature ΔT in this phase increases linearly with the absorbed heat ΔQ as $\Delta T = \Delta Q / c_p m$, where $c_p$ is the specific heat and $m$ is the mass of the excited VO$_2$ film. The temperature of the film increases more in the tetragonal phase than during the phase transition for the same increase of pump fluence since part of the absorbed energy was supplied as the latent heat during the phase transition. For a fluence higher than 14.7 mJ/cm$^2$, the temperature of the film after photoexcitation can be corrected by a scale of $C_{IM}/C_T$=2.43, where $C_{IM}$ is the heat capacity of the film during the transition as calculated in Sec. III and $C_T$ is the heat capacity in the tetragonal phase listed in Table 1. Similar rescaling with the heat capacity in the monoclinic phase can be applied to the fluence range below 7 mJ/cm$^2$. Therefore, the film temperature was calibrated at 100 ps after photoexcitation for a given optical pump fluence.

When the pump fluence was higher than 14.7 mJ/cm$^2$, the lattice response started to deviate from the thermally induced changes by prominent effects on Bragg peak position and intensity [Fig.2 (b), (c)]. The Bragg peak moved to an even lower angle and the diffraction intensity was smaller than that of the tetragonal phase. Based on the fluence-temperature calibration, the film temperature is 552 K at 100 ps when pumped by a fluence of 29 mJ/cm$^2$. This corresponds to an additional 0.1% expansion of the out-of-plane lattice of tetragonal VO$_2$ using a thermal expansion coefficient of $\alpha = 4.87 \times 10^{-6}$ K$^{-1}$ [40]. Correspondingly, the Bragg peak position would be further moved to lower angle by 0.034°, which is well under the observed 0.08° shift. The upper bound of the film temperature is calculated to be 669 K, under the assumption that all absorbed optical energy is converted to heat. This is higher than the calibrated temperature of 552 K due to possible energy relaxation pathways which do not heat the film, such as photoluminescence [41]. Even at 669 K, the expected thermal expansion is still smaller than observed lattice expansion. In addition, because the VO$_2$ lattice was incommensurate with the substrate, in-plane stress is not a major contribution to out-of-plane strain. Thus, we can exclude the Poisson's effect as a significant source of lattice expansion, although it may play a role in commensurate films [42]. The diffraction intensity of the Bragg peak excited by high fluence laser pulses was reduced lower than that could be expected from the Debye-Waller effect as a result of increased temperature. Based on these experimental evidence, the film at 100 ps after high fluence photoexcitation corresponds to a new thermally inaccessible structural state. This structural state exhibited distinct recovery pathway rather than thermal cooling as discussed in the following section.

d. **The recovery dynamics of photoexcited VO$_2$ thin film**

To study the structural recovery dynamics, we monitored the 400 Bragg reflections of the VO$_2$ film at various time delays after laser excitation. The peak positions and intensities were extracted and plotted as a function of the delay in Fig. 3(a) and (b) respectively. The recovery dynamics of the Bragg peak was highly dependent on the pump fluence, similar to the observation of the electronic recovery of photoexcited $VO_2$ [30]. At pump fluence lower than 14.7 mJ/cm$^2$, photoexcitation was not sufficient to completely transform the film from the monoclinic phase to tetragonal phase. For example, at a fluence of 11.7 mJ/cm$^2$, the $VO_2$ film was excited to an intermediate state as marked by the blue star in Fig. 2(c) and recovered following the temperature-dependent thermal pathway. The position of the Bragg peak recovered in 2 ns while the intensity recovered in 10 ns to the 1/$e$ values [Fig. 3(a)(b)]. At fluences higher than 14.7 mJ/cm$^2$, the Bragg peak shifted to a lower angle beyond the position of the Bragg peak of the tetragonal lattice, consistent with a new structural state as seen in the high-fluence regime in Fig. 2(c). For example, at a fluence of 29 mJ/cm$^2$, the out-of-plane expansion of the lattice reverted to the lattice parameter of the tetragonal phase with a time constant of 700 ps, which is faster than the expected thermal recovery time. The Bragg peak position remained constant until the film temperature cooled to the transition temperature. Subsequently, the film recovered to the monoclinic phase with a characteristic time of several to tens of ns as a result of a thermal transport process discussed in Sec.III. The diffraction intensity decreased within 100 ps and then sharply increased to higher intensity, followed by a slow recovery in tens of ns. The initial reduction of the diffraction intensity within 100 ps indicates the out-of-plane lattice disorder, consistent with a recent report [43].

To further investigate this new transient state, we repeated the time-resolved diffraction at the elevated film temperature of 375 K, where the $VO_2$ film was already in the tetragonal phase. At a fluence of 15.2 mJ/cm$^2$, the position of the Bragg reflection decreased upon photoexcitation beyond the tetragonal position and recovered at the same time constant of 700 ps as the fast recovery of photoexcited monoclinic $VO_2$ excited by a fluence of 29 mJ/cm$^2$ [Fig. 3(d)]. The agreement of these two recovery time scales suggests that the 700-ps recovery in photoexcited monoclinic $VO_2$ shares the same origin as the structural change in photoexcited rutile $VO_2$. The photoexcitation of monoclinic $VO_2$ by a fluence higher than the saturation fluence can be understood as a two-step process. The first step is the photoinduced phase transition that consumes the incident optical energy 14.7 mJ/cm$^2$, i.e., the fluence which completely transformed the film to the tetragonal phase at 100 ps. The excessive incident optical energy higher than 14.7 mJ/cm$^2$ is responsible to excite $VO_2$ film from the tetragonal phase to a new structural state with a larger out-of-plane lattice parameter. Since the recovery time is independent of electronic properties of $VO_2$ before the excitation, this transient structural state may not be related to the electronic phase transition, as supported by the electronic recovery dynamics discussed as follows.

To study the electronic recovery of the $VO_2$ films, we performed transient optical absorption measurements at the Center for Nanoscale Materials, Argonne National Laboratory. After excited by a 40-fs, 800-nm pulse, $VO_2$ film was probed by a chirped 1 ps infrared broadband pulse in the wavelength range from 850 to 1600 nm with time delays up to 7.2 ns. The transmission broadly reduced across the probe spectrum with the maximum change around 1450 nm as a result of the change of the film conductivity. Since the change of induced absorption did not depend on the probe wavelength, the time-dependent absorption at 1450 nm was used to measure the change of the optical conductivity of the film as a function of delay. Comparing with the structural dynamics, we found that the time scales for recovery of the transmission largely agreed with the recovery of the Bragg peak position (Fig. 4) but did not match the recovery of the Bragg peak intensity, supporting the hypothesis that the electronic phase transition is sensitively related to the lattice structure but not the disorder of the lattice. More importantly, we note the transient structural state that recovers in 700 ps was not observed in the recovery of the induced absorption. Therefore, the new structural state observed at high pump fluence was not accompanied by a change in the electronic properties and independent of the photoinduced MIT.

Although the experimental evidence supports this new transient structural state at 100 ps is non-thermally excited by an intense optical pulse, the excitation mechanism still needs to be understood. High carrier concentrations upon intense optical excitation can alter the potential energy surface of the lattice and reduce Bragg peak intensity in photoexcited bismuth [44]. Similarly, highly excited $VO_2$ also shows the diffraction intensity reduction larger than thermally expected values at ps time scales [43]. The time scale from several ps to ns is a cross-over region where the direct electronic contribution to the lattice change gives way to thermal effects. Our work provides the evidence that the lattice change at 100 ps, in the cross-over time scale, may not only arise from the temperature increase of the film but also be related to electronic contributions such as bond softening due to deformation of potential surface of the lattice upon intense photoexcitation [44]. Further experiments and theory are needed to understand the nature of this new transient structural state of $VO_2$.

### III. Thermal transport model and simulation

To quantitatively understand the recovery dynamics, we simulated the change of the Bragg reflection as a function of time by solving a one-dimensional thermal diffusion equation for a heterostructure [45]. The heat transport after the initial excitation can be described by

$$\frac{dT(t,z)}{dt} = \left(\frac{\kappa}{C_p \rho}\right) \frac{d^2 T(t,z)}{dz^2} \tag{1}$$

where T is the temperature profile as a function of time *t* and depth *z* from the $VO_2$-air interface into the sample. The boundary conditions at the interface are described by

$$C_p\rho \frac{dT(t,z)}{dt} = -g\frac{dT(t,z)}{dz}|_{z=D}, \frac{dT(t,z)}{dz}|_{z=0} = 0, \quad (2)$$

and the initial condition is

$$T(0 < z < D, t = 0) = T_0 \exp(-\alpha z). \quad (3)$$

Here, $T_0$ is the initial surface temperature, $\alpha$, $g$, $D$, $C_p$, $\rho$, and $\kappa$ are material parameters listed in Table 1. The average film temperature at any instant is given by $T(t) = \frac{1}{D}\int_0^D T(t,z)dz$. Transverse heat transfer can be ignored because the excitation area is much larger than the area probed by the x-ray beam. The thermal transport between air and sample is negligible at ns time scales. During the first-order phase transition, the heat capacity has a higher value due to the absorption of latent heat [12]. We approximated the empirical heat capacity during the phase transition by $C_{IM} = (C_M + C_T)/2 + L/\Delta T$, where $C_M$ and $C_T$ are the heat capacity of monoclinic and tetragonal VO$_2$ respectively. $L$ is the latent heat and $\Delta T = (T_2 - T_1)$ is the temperature range where the structural transition starts to occur at the temperature of $T_1$ and completes at $T_2$ as displayed in Fig. 2(c).

Using the parameters of the VO$_2$ film and TiO$_2$ substrate listed in Table 1, we calculated the temperature of the VO$_2$ film as a function of time based on the initial temperature calibrated in Fig.2(c) and display the results in the insets of Fig. 5. Although the heat transfer was mainly determined by the thermal properties of the film and the substrate, the characteristic time of the thermal recovery across the transition was highly dependent on the incident fluence. The increase of the pump fluence yielded higher initial temperature after photoexcitation and higher base temperature before photoexcitation. These two effects can significantly change the cooling rate across the phase transition. For example, at the pump fluence of 14.3 mJ/cm$^2$, the initial temperature of the film after photoexcitation was derived based on Fig. 2(c) to be 359 K. Since the substrate remains unheated when the phase transition occurs, the large temperature difference between the film and substrate produces a fast change of the film temperature and fast recovery across the phase transition [Fig. 5(a), inset]. At the pump fluence of 29 mJ/cm$^2$, the initial temperature of the film was driven to 552 K. As the film cooled down to the transition temperature, heating of the near-interface substrate reduced the temperature difference between the film and the substrate thus lowering the cooling rate. As a result, the temperature decreased more slowly across the phase transition [Fig. 5(b), inset].

Once the film temperature as a function of time is obtained, we can simulate the temporal change of the Bragg peak after photoexcitation. For any given time, the corresponding Bragg peak position and intensity can be retrieved from Fig. 2(c) after correcting for hysteresis during the cooling of the film. At a pump fluence of 14.3 mJ/cm$^2$, the simulated results agree with the measurement well [Fig. 5(a)]. At a pump fluence of 29 mJ/cm$^2$, the thermal model predicts most of the observed features except for the large initial (<1ns) shift of the Bragg peak to lower angle beyond the tetragonal phase [Fig. 5(b)]. This confirms that the non-thermally induced transient structural state as observed at high pump fluence follows a fast relaxation of possible photoinduced lattice softening rather than a thermal cooling pathway.

This one-dimensional thermal transport model successfully predicts the time-dependent thermal recovery pathways and the strong dependence of the recovery process on the pump fluence. It provides a framework to the thermal transport modeling of devices based on photoexcited VO$_2$ films.

## IV. Conclusion and discussion

We have studied the structural and electronic recovery dynamics of an ultrathin VO$_2$ film after photoexcitation on ns time scales. The VO$_2$ thin film can be excited to a selected structural state by optical pulses. At high pump fluences, we observed a new transient structural state with a larger lattice parameter than that of the tetragonal phase, which neither is thermally excited nor follows a thermally predicted pathway and is decoupled from the metal-to-insulator phase transition. The subsequent recovery from the tetragonal metallic phase to the monoclinic insulating phase is mainly thermally driven and can be controlled by the pump fluence. The control mechanism is understood as a result of the different cooling rates across the transition, as predicted by a thermal transport model that explains the fluence-dependent recovery pathways of photoexcited $VO_2$. The understanding of the recovery dynamics suggests the reversible switching of $VO_2$ films between the metallic and insulating phases may achieve higher rates by using thinner film and optimizing interfacial thermal transport in a $VO_2$ heterostructure.

**Acknowledgments:**

Work at Argonne was supported by the U.S Department of Energy, Office of Science, Office of Basic Energy Sciences, under Contract No. DE-AC02-06CH11357. V.G. would like to acknowledge funding from the Office of Naval Research (ONR) award number N00014-11-1-0665. Work at Cornell University was supported by Army Research Office through agreement W911NF-08-2-0032. The transport measurements were performed in Peter Schiffer's laboratory.

**Figure Caption:**

Fig. 1 The resistivity of the $VO_2$ film as a function of temperature. The inset shows the monoclinic and tetragonal lattice structures of $VO_2$.

Fig. 2 a) Representative 400 Bragg peaks at various temperatures. "M" and "T" mark the peak position of monoclinic and tetragonal lattice structures of $VO_2$. b) Representative 400 Bragg peaks at 100 ps after 800nm optical excitation at various pump fluences. The counting statistics is worse than that in a) because the data-collection rate is significantly reduced to match the repetition rate of the pump laser. c) The change of 400 Bragg peak position (square) and intensity (circle) as a function of the film temperature (black) or pump fluence (red) at 100 ps after photoexcitation. The stars represent the initial states after photoexcitation at specific fluences, with corresponding arrows showing the recovery pathways of the Bragg peak (see text). The solid vertical lines indicate the threshold and saturation fluences of the structural phase transition with the corresponding temperature $T_1$ and $T_2$, which divide the plot into three regions. The light blue region indicates the monoclinic structure phase. The white region indicates the fluence or temperature range where the phase transition occurs. The light yellow region indicates the fluence range where non-thermal effects are prominent upon photoexcitation. The dashed lines show the projected thermal recovery pathways.

Fig. 3 a) The change of 400 Bragg peak position as a function of the pump-probe delay at various fluences. The dashed lines indicate the Bragg peak position of monoclinic and tetragonal $VO_2$. b) The change of 400 Bragg peak intensity as a function of the pump-probe delay at various fluences. c) The change of 400 Bragg peak position as a function of the pump-probe delay at the base sample temperature below (313 K, triangles) and above (375 K, diamonds) the transition temperature.

Fig. 4 The change of lattice parameter (symbols/dashed lines) measured by ultrafast x-ray diffraction and the induced absorption (OD: optical density) at 1450 nm (thick lines) measured by ultrafast transition absorption spectroscopy as a function of the delay.

Fig. 5 The change of 400 Bragg peak position and integrated intensity (circles) as a function of the pump-probe delay, compared with a thermal transport model (solid lines) at pump fluences of (a) 14.3 mJ/cm$^2$ and (b) 29 mJ/cm$^2$. The insets show the calculated film temperature as a function of time. The dashed lines indicate the temperature range where the transition occurs.

**Figure 1, Wen et. al.**

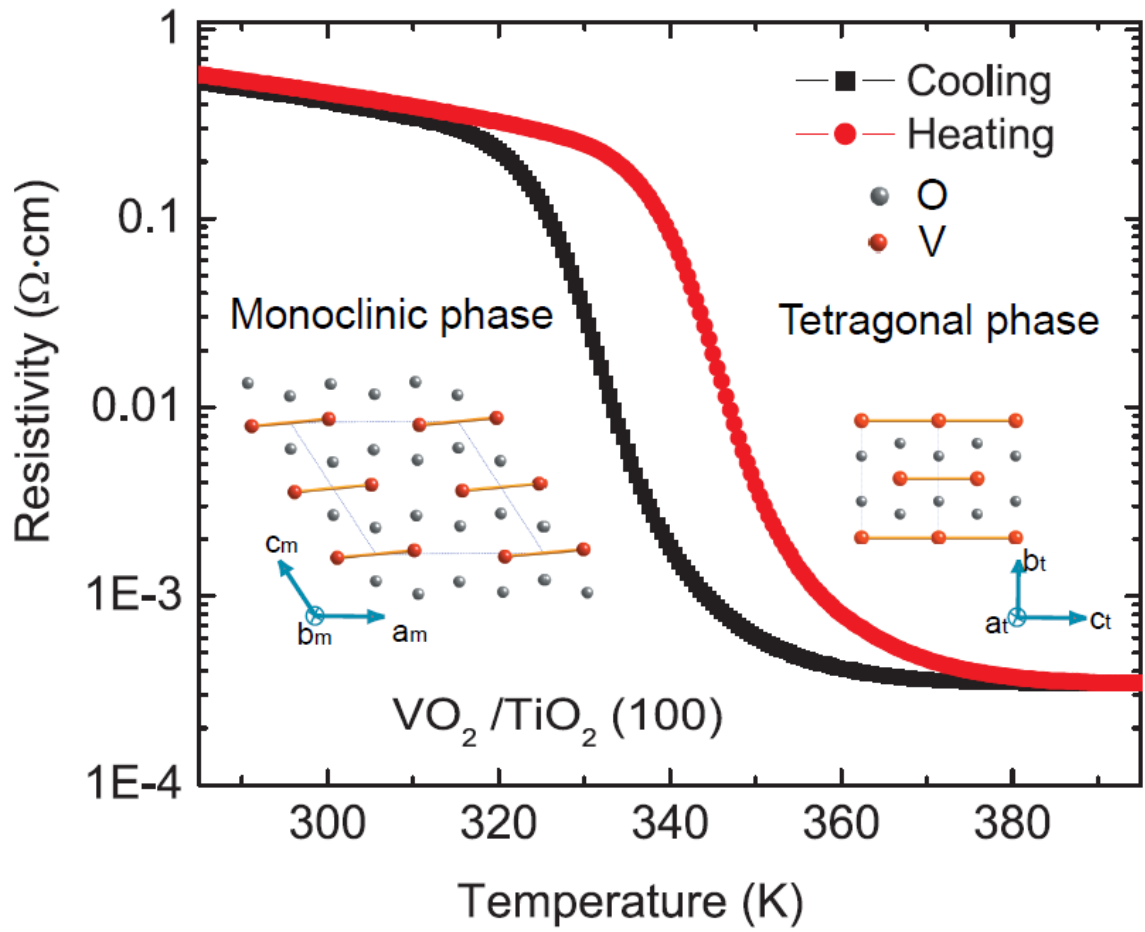

**Figure 2, Wen, et. al.**

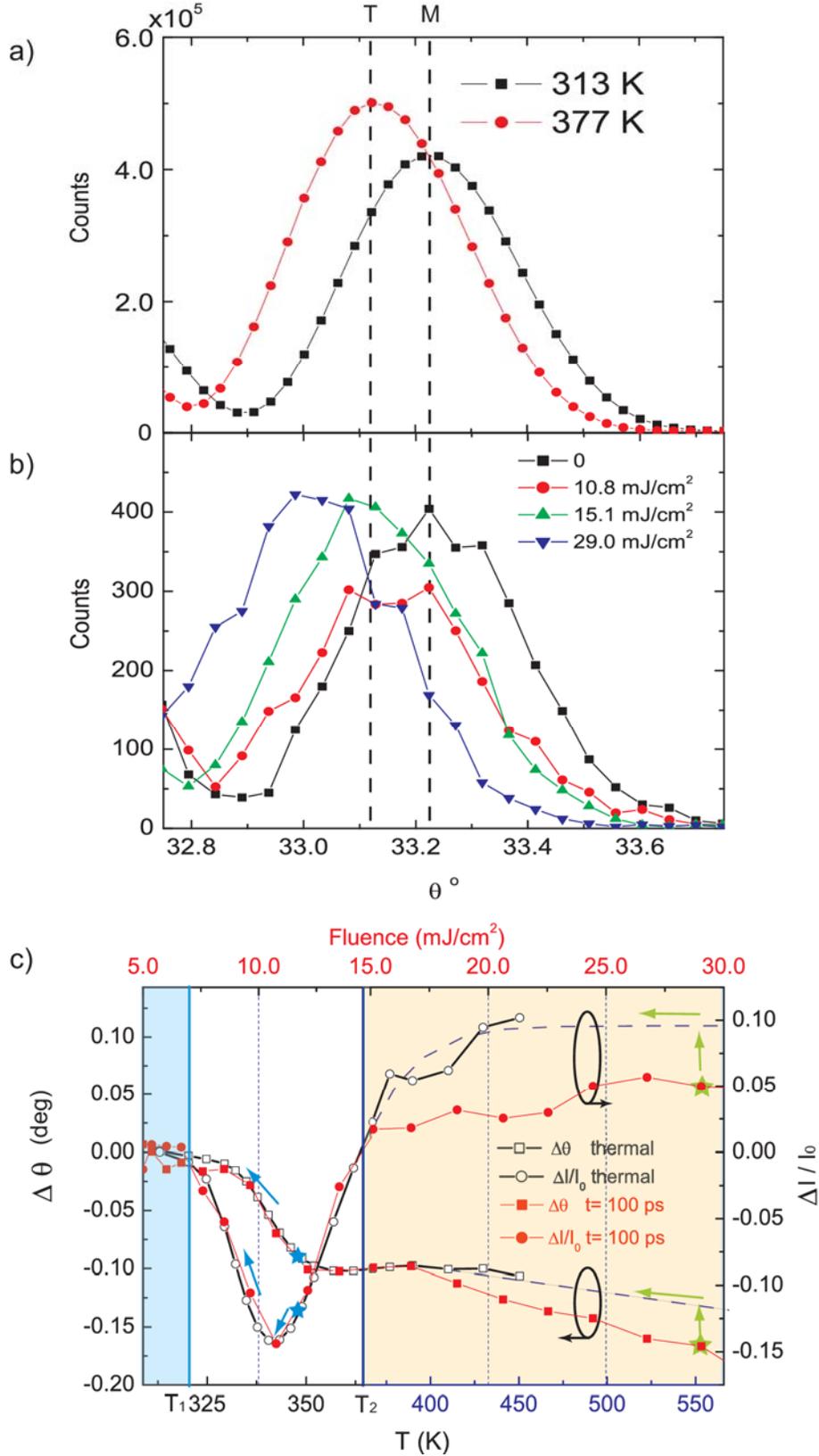

**Figure 3, Wen, et. al.**

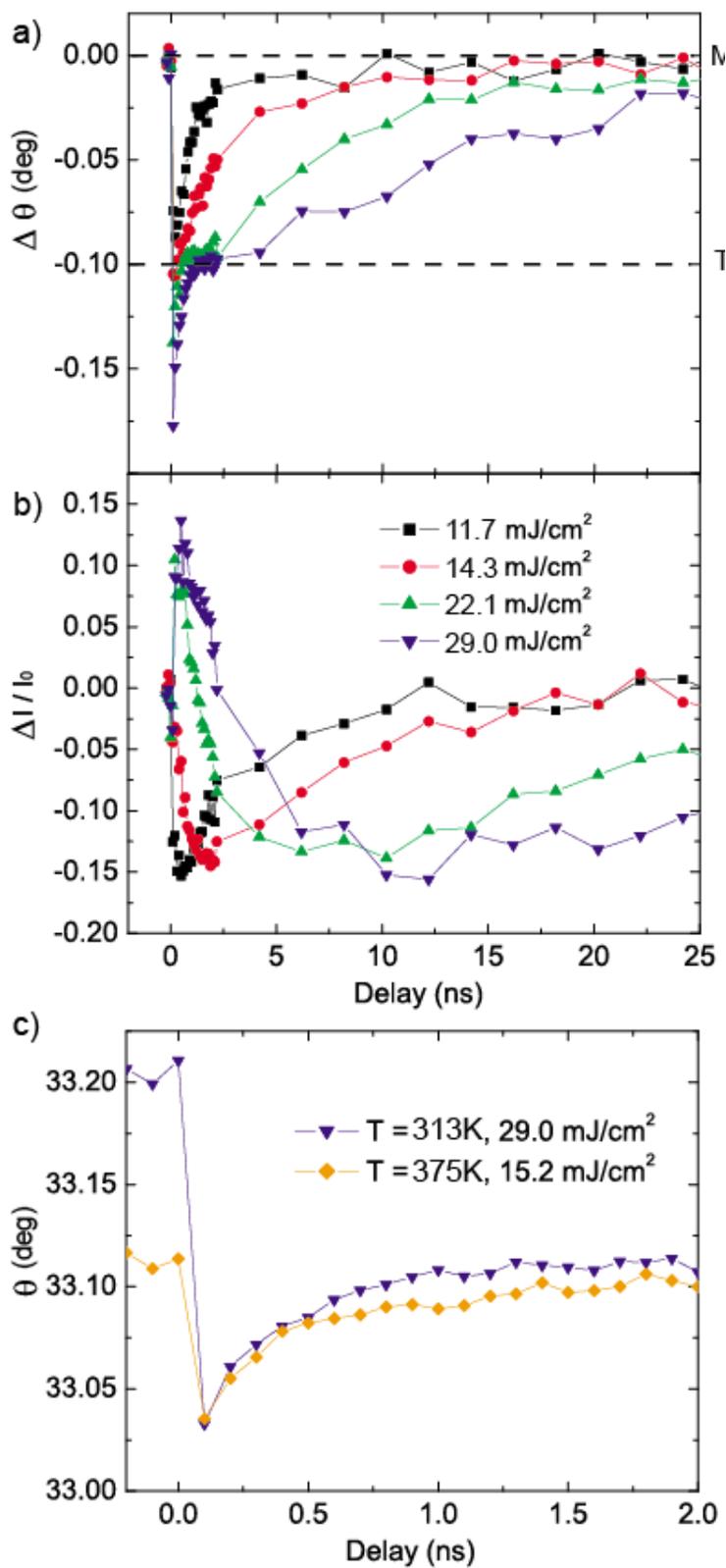

**Figure 4, Wen, et. al.**

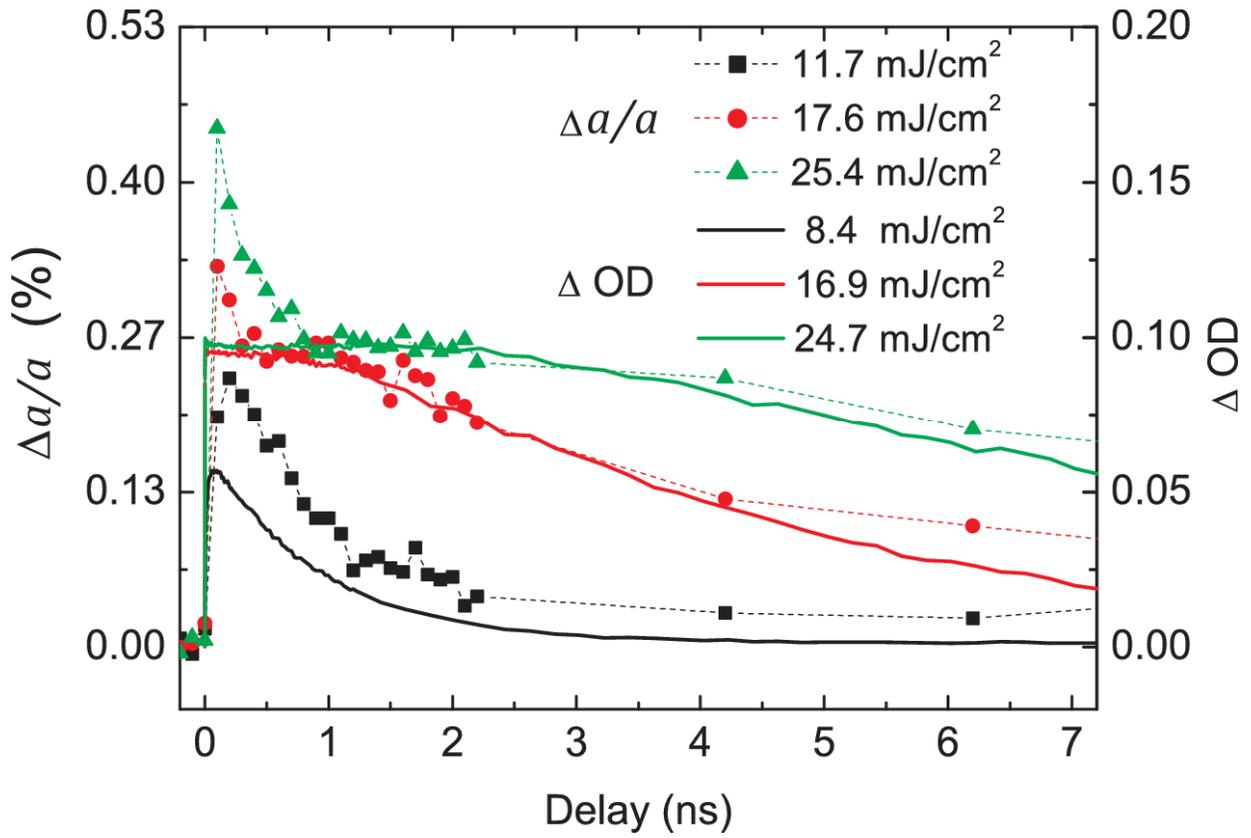

**Figure 5, Wen, et. al.**

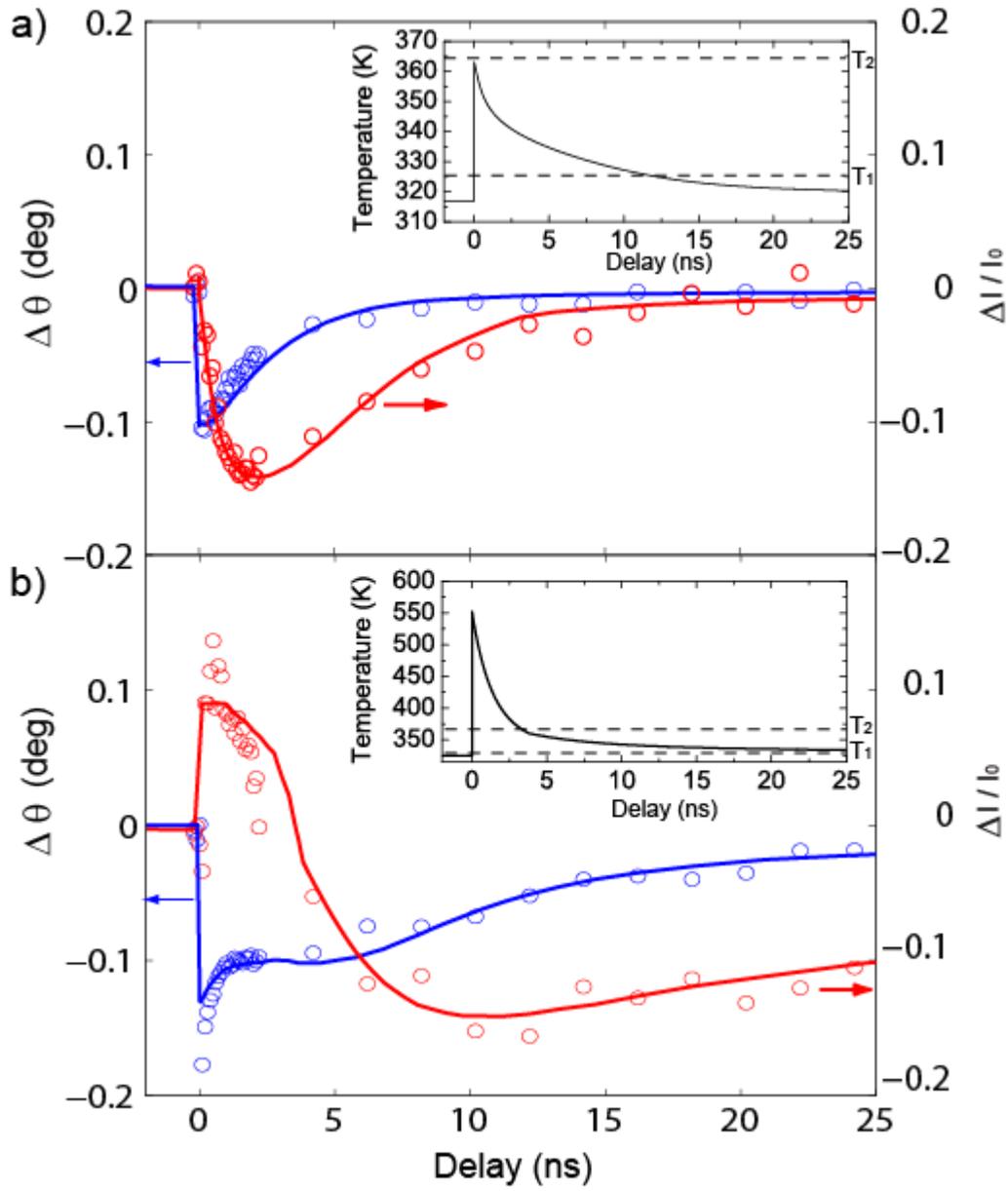

**Table 1: Parameters of VO$_2$ and TiO$_2$**

| | |
|---|---|
| VO$_2$ absorption coefficient for 800nm light $\alpha$ [39] | 0.01 nm$^{-1}$ |
| VO$_2$ film thickness $D$ [a] | 13 nm |
| VO$_2$ density $\rho_1$ [b] | Monoclinic: 4.57 g/cm$^3$<br>Tetragonal: 4.65 g/cm$^3$ |
| TiO$_2$ density $\rho_2$ [b] | 4.25 g/cm$^3$ |
| VO$_2$ thermal expansion coefficient α [40] | 4.87×10$^{-6}$/K |
| VO$_2$ thermal conductivity $\kappa_1$ [44] | Monoclinic 3.5 W/(m·K)<br>Tetragonal: 6 W/(m·K) |
| TiO$_2$ thermal conductivity $\kappa_2$ [47] | 8 W/(m·K) |
| Kaptiza (interfacial) conductance $g$ [c] | 3300 W/(K·cm$^2$) |
| VO$_2$ heat capacity $C_{1p}$ [12] | Monoclinic: 0.656 J/(g·K)<br>Tetragonal: 0.78 J/(g·K) |
| TiO$_2$ heat capacity $C_{2p}$ at 300 K [48] | 0.686 J/(g·K) |
| VO$_2$ latent heat $L$ [12] | 51.8 J/g |
| The lower bound of the phase transition temperature $T_1$ [a] | 322 K |
| The upper bound of the phase transition temperature $T_2$ [a] | 366 K |

[a] Measured value; [b] Calculated value based on atomic mass and unit cell volume; [c] Fitting result.